\documentclass[baaa]{baaa}

\usepackage[pdftex]{hyperref}
\usepackage{subfigure}
\usepackage{natbib}
\usepackage{helvet,soul}
\usepackage[font=small]{caption}
\usepackage{xcolor}
\usepackage[normalem]{ulem}

\contriblanguage{1}

\contribtype{1}

\thematicarea{7}

\received{\ldots}
\accepted{\ldots}

\title{Strong gravitational lensing constraints on interacting dark energy models}

\titlerunning{Strong gravitational lensing constraints on interacting dark energy models}

\author{
F. Villalobos\inst{1},
T. Verdugo\inst{2},
P. Troncoso-Iribarren\inst{1},
N. Piña\inst{1}
\&
J. Magaña\inst{1}
}

\authorrunning{Villalobos et al.}

\contact{felipevillalobos453@gmail.com}

\institute{
Facultad de Ingeniería y Arquitectura, Universidad Central de Chile
\and
Instituto de Astronom\'ia, Observatorio Astron\'omico Nacional, Universidad Nacional Aut\'onoma de M\'exico, Apartado postal 106, C.P. 22800,  Ensenada, B.C., M\'exico.
}

\resumen{La posible interacción entre la materia oscura y la energía oscura ha sido propuesta como un mecanismo para aliviar el problema de la coincidencia y la tensión de Hubble. Las observaciones de lentes gravitacionales fuertes proporcionan valiosas restricciones para probar las propiedades de los componentes oscuros del Universo. Presentamos la estimación de los parámetros cosmológicos empleando datos de lentes gravitacionales fuertes para un modelo de interacción $Q$, proporcional a la densidad de materia oscura y dependiente del parámetro de desaceleración $q$. Los resultados obtenidos son consistentes con un Universo en expansión acelerada. Además, concuerdan con estudios previos que sugieren que una interacción en el sector oscuro podría contribuir a aliviar las tensiones existentes en la historia de expansión del Universo y resaltan el potencial de las lentes gravitacionales fuertes como una sonda cosmológica independiente.}

\abstract{
The possible interaction between dark matter and dark energy has been proposed as a mechanism to alleviate the coincidence problem and the Hubble tension. Strong gravitational lensing observations provide valuable constraints to test the properties of the dark components of the Universe. We present estimates of cosmological parameters employing strong lensing data for an interaction model $Q$, proportional to the dark matter density and dependent on the deceleration parameter $q$. The obtained results are consistent with an accelerating Universe. Furthermore, these results agree with previous studies suggesting that an interaction in the dark sector may help alleviate existing tensions in the expansion history of the Universe and highlight the potential of strong gravitational lensing as an independent cosmological probe.}


\begin{document}

\maketitle
\section{Introduction}\label{S_intro}

In the current standard cosmological model within general relativity, it is well known that there is evidence for the accelerated expansion of the Universe sourced by a repulsive dark energy (DE) component in the form of a cosmological constant \citep{riess_SN1a_1998, perlmutter_SN1a_1999}. On the other hand, the model also postulates the existence of a cold dark matter component responsible for the formation of large-scale structure.

In this model, known as the $\Lambda$ cold dark matter ($\Lambda$CDM) model, dark matter (DM) accounts for approximately $27\%$ of the total energy content of the Universe, while DE contributes about $68\%$; the remainder corresponds to baryonic matter. Although this paradigm is strongly supported by several cosmological observations such as the cosmic microwave background (CMB) radiation \citep{Planck_2018} and Type Ia supernovae (SNeIa) \citep{Scolnic:2021amr}, as well as cosmic chronometers \citep{moresco_CC_2016}, among others, it still presents several open issues. For instance, the so-called \emph{coincidence problem} \citep{Copeland2006, Bamba2012} which arises from the near equality of the present-day energy densities of DM and DE, despite their distinct physical nature and evolutionary histories. Another challenge faced by $\Lambda$CDM is the Hubble tension \citep{DiValentino2021}, which refers to the discrepancy between the estimations of the so-called Hubble constant ($H_0$) inferred from late and early universe data. There are rigorous methods used to measure $H_0$, which represents the expansion rate of the Universe. Among the most precise measurements are those using SNeIa, which act as standard candles by relating their intrinsic luminosity to the observed distance \citep{riess_SN1a_1998, perlmutter_SN1a_1999}. Additionally, there is an alternative approach that does not use local object measurements, which is based on the CMB radiation that allows inferring $H_0$ from early Universe anisotropies \citep{Planck_2018}.
While \citet{Planck_2018} report a value of $H_0=67.37\pm0.54$~km/s/Mpc, the late data reports $H_0=73.30\pm1.04$~km/s/Mpc \citep{Riess:2022ApJ}, indicating a tension of $5\sigma$ which is not explained within the $\Lambda$CDM model.

To address these issues, a wide variety of alternative models has been proposed, including dynamical DE scenarios based on specific parametrizations, as well as modifications of gravity (see \citealt{Motta:2021hvl} for a review). Another interesting possibility is that DM and DE interact with each other, i.e there is an energy transfer from one dark component to the other through an interacting term $Q$.
These interacting dark energy (IDE) scenarios (see \citealt{Bolotin:2013jpa} for a review) offer a plausible mechanism to explain why the densities of both components are comparable today. In addition, they have been proposed as a possible way to alleviate the Hubble tension \citep{diValentino:2017}.
An interesting kind of IDE model is the one where there is a change of direction in the energy transfer, i.e. a change in sign in the dark interaction, during the cosmic evolution.

In the present work, following the approach proposed by \citet{verdugo_EPJC_2024}, we aim to test an interacting dark sector model characterized by a sign-changeable coupling function $Q$. To this end, we combine two complementary strong-lensing constraints operating at different physical scales: the mass distribution of the well-known galaxy cluster Abell~1689 \citep{jullo_SL_2010}, and the lensing properties of early-type galaxies acting as strong lenses \citep{amante_SL_2020}. 
The outline of the manuscript is as follows: Sec. \ref{sec:IDE model} is dedicated to presenting the mathematical formalism of sign-changeable IDE models. Sec. \ref{sec:methodology} describes the data and methodology used to constrain the IDE model. In Sec. \ref{sec:Results}, we show our results and finally in Sec. \ref{sec:Conclusions} we give our conclusions.

\section{The interacting Q model}
\label{sec:IDE model}

In this work, we present parameter estimations for an interacting $Q$ model proportional to the DM density and dependent on the deceleration parameter $q$. Such models are motivated by observational indications of a possible change in the interaction term within the range $0.45 \lesssim z \lesssim 0.9$ \citep{Cai_Su_2010}. Inspired by these results, \citet{wei_interaction_2011} introduced a phenomenological interacting dark-energy model in which the interaction term depends on the deceleration parameter $q$. In this framework, the cosmological dynamics are governed by
\begin{eqnarray}
\dot{\rho}_{dm}+3H\rho_{dm} &=& Q, \nonumber\\
\dot{\rho}_{de}+3H(\rho_{de}+p_{de}) &=& -Q,
\end{eqnarray}
where $H$ is the Hubble parameter, while $\rho_{dm}$ and $\rho_{de}$ are the DM and DE densities, respectively. Following \citet{wei_interaction_2011}, the interaction term is given by

\begin{equation}
Q = 3\beta H \rho_{dm} q,
\end{equation}
where $\beta$ is a dimensionless coupling constant.

With this choice, the dimensionless Friedmann equation becomes

\begin{align}
E &\equiv \frac{H}{H_0} \nonumber\\
  &= \left[ 1 - \frac{2 - 3\beta}{2(1+\beta)}\,\Omega_{dm0}
     \left( 1 - (1+z)^{3(1+\beta)} \right)
     \right]^{\frac{1}{2+3\beta}},
     \label{Friedmann}
\end{align}
where $\Omega_{dm0}$ is the DM density parameter at $z = 0$. Thus, the model contains two free parameters, $\left(\Omega_{dm0},\beta\right)$, and the standard cosmological model is recovered when $\beta=0$. The deceleration parameter reads

\begin{multline}\label{eq:qI}
q(z) = \frac{1+z}{2\,E(z)^2}\,\Bigl\{ 3\,\Omega_{dm0}\,(1+z)^{3(1+\beta)-1} \\
\times \Bigl( 1 - \frac{2+3\beta}{2(1+\beta)}\,\Omega_{dm0}
\bigl[ 1 - (1+z)^{3(1+\beta)} \bigr] \Bigr)^{\frac{2}{2+3\beta}-1} \Bigr\} - 1 .
\end{multline}
where $E(z)$ is given by Eq.\eqref{Friedmann}.
\section{Data and methodology}\label{sec:methodology}

We employ the compilation of early-type galaxies acting as strong gravitational lenses presented by \citet{amante_SL_2020}. The fiducial sample comprises \( N_{\mathrm{SL}} = 143 \) strong lensing systems (SLS), each characterized by four measured quantities: the stellar velocity dispersion (\( \sigma \)), the Einstein radius (\( \theta_E \)), the lens redshift (\( z_l \)), and the source redshift (\( z_s \)).

We can constrain cosmological parameters minimizing the chi square function given as
\begin{equation}
\chi_{\mbox{gal}}^2 = \sum_{i=1}^{N_{SL}} \frac{ \left[ D^{th}\left(z_{l}, z_{s}; \bf{\Theta_{Cos}} \right)  -D^{obs}(\theta_{E},\sigma^2)\right]^2 }{ (\delta D^{\rm{obs}})^2},
\label{eq:chisquareSL}
\end{equation}
where we define the ratio of two angular diameter distances $D \equiv D_{ls}/D_{s}$. The theoretical ratio $D^{th}$ is calculated from Eq.\,\ref{Friedmann}, using the definition of angular diameter distance. On the other hand, through the observationally measured properties, we get the observed counterpart $D^{obs}$ assuming a singular isothermal sphere (SIS) model for the lens galaxy, where $D^{obs} = \frac{c^2 \theta_E}{4\pi \sigma^2}$ (see e.g. \citealt{amante_SL_2020}), and $\delta D^{\rm{obs}}$  as the error propagation of the $D^{obs}$ function. The vector $\mathbf{\Theta_{Cos}}$ is formed by the parameters $\beta$ and $\Omega_{dm0}$.

\begin{figure}[!t]
\centering
\includegraphics[width=\columnwidth]{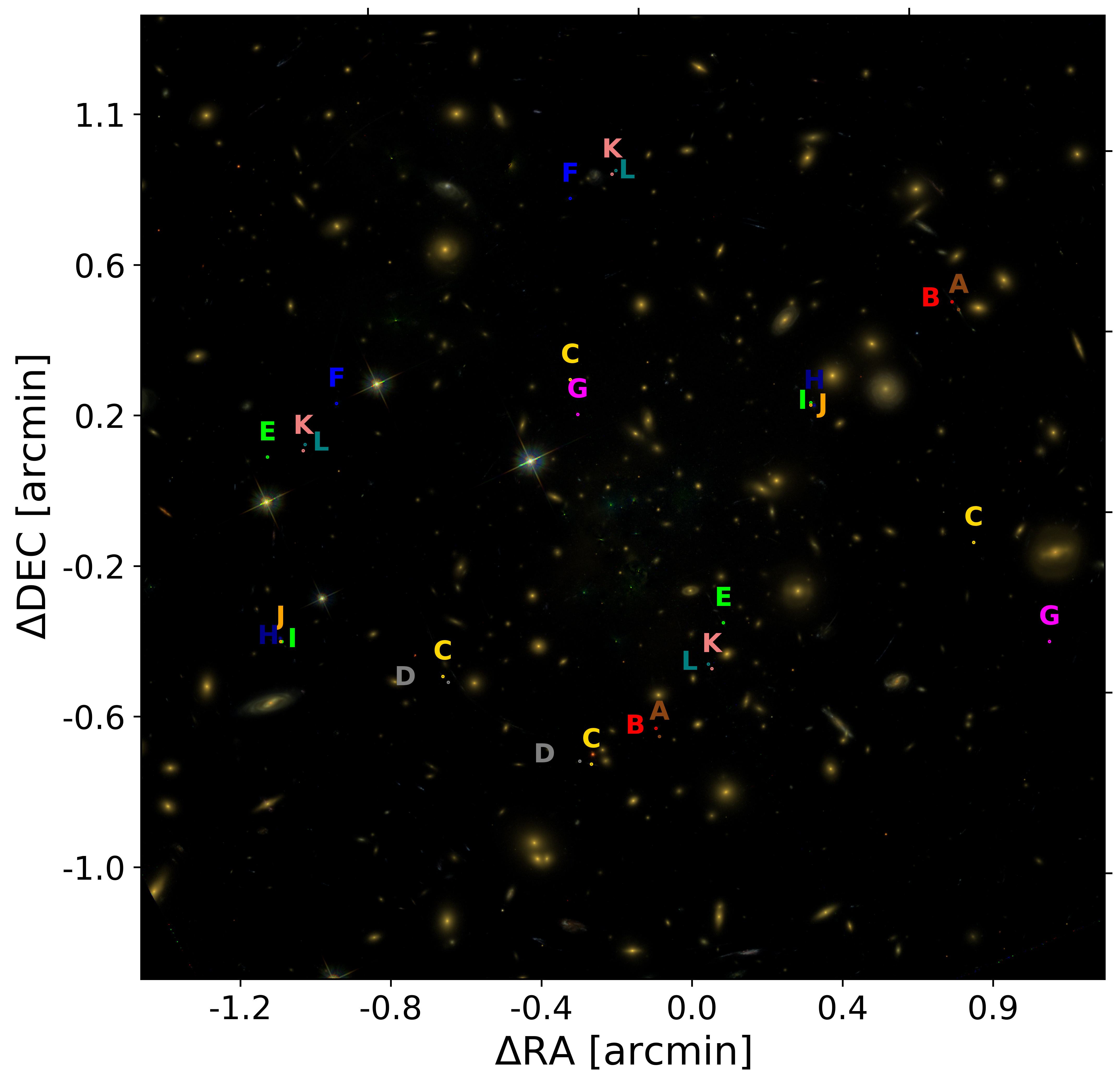}
\caption{HST/ACS colour composite image (F475W, F625W, and F775W) of Abell 1689 based on archival observations from HST program 9289, available through the Mikulski Archive for Space Telescopes \citep[see][]{2007ApJ...668..643L, 2005ApJ...621...53B}. Labels indicate the multiple-image systems used in the model. Different colors represent distinct families of multiple images, with circles highlighting the 28 images from 12 families.}
\label{fig:abell1689}
\end{figure}




\subsection{Strong lensing in Abell 1689}

Following \cite{jullo_SL_2010}, to reconstruct the A1689 mass model and simultaneously  constrain the cosmological parameters  we use the ``family ratio'' which is defined  as the angular diameter distance ratios of two images from different sources. Our model uses 28 images from $N_f$ = 12 families (see Fig.~\ref{fig:abell1689}), all with measured spectroscopic redshifts in the range $1.15< z_{\rm s} < 4.86$. For the present work, we adopt an error $\Delta^2$ = 0.5$^{\prime\prime}$ in the position of the images. In this case, the $\chi^{2}$ for a multiple image system $i$ is defined as 
\begin{equation}\label{eq:Chi2Lens}
\chi_{i}^{2} = \sum_{j=1}^{n_i}
\frac{\left| \vec{x}_{\rm obs}^j - \vec{x}^j(\mathbf{\Theta_{W}}) \right|^2}{\Delta^{2}}\;,
\end{equation}
where $n_i$ is the number of multiple images for the source $i$, $\vec{x}_{\rm obs}^j$ is the observed position corresponding to image $j$, and $\vec{x}^j(\mathbf{\Theta_{W}})$ is the position of image $j$ predicted by the  model, whose total parameters (the cosmological parameters and the cluster parameters) are included in the
vector $\mathbf{\Theta_{W}}$. Thus, $\chi^2_{\mbox{clu}} =\sum_{1}^{N_f} \chi_{i}^{2}$, is the total chi square function in this case.

\subsection{Cosmological parameter estimation}

Assuming a Gaussian likelihood, $\mathcal{L} \propto e^{-\chi^{2}_{\mathrm{tot}}/2}$, the total chi-square function is defined as
\begin{equation}\label{eq:Chi2Tot}
\chi_{\mathrm{tot}}^{2} = \chi^2_{\mathrm{gal}} + \chi^2_{\mathrm{clu}},
\end{equation}
where $\chi^{2}_{\mathrm{tot}}$ represents the combined contribution of both complementary strong-lensing approaches: the sample of early-type galaxies acting as lenses and the cluster-scale constraints from Abell~1689. This total chi-square is minimized within {\sc LENSTOOL} using a Bayesian Markov Chain Monte Carlo algorithm, following the methodology described in \citet{verdugo_EPJC_2024}, to obtain the best-fit cosmological model parameters.

Our Bayesian analysis employs a $3\sigma$ Gaussian prior on the matter density parameter 
$\Omega_{m0} = 0.311 \pm 0.006$ following \citet{Planck_2018}, 
and a uniform prior $\beta \in [-1,1]$ for the interacting parameter. 
This range avoids regions where the expansion rate becomes non-real or numerically unstable, particularly for $\beta \lesssim -1$, and ensures a well-defined likelihood over the explored parameter space.
Extending the prior toward more negative values does not improve the fit, as these regions are associated with unphysical or ill-defined cosmological solutions.




\section{Results and Discussion}\label{sec:Results}

\begin{figure}[!t]
\centering
\includegraphics[width=\columnwidth]{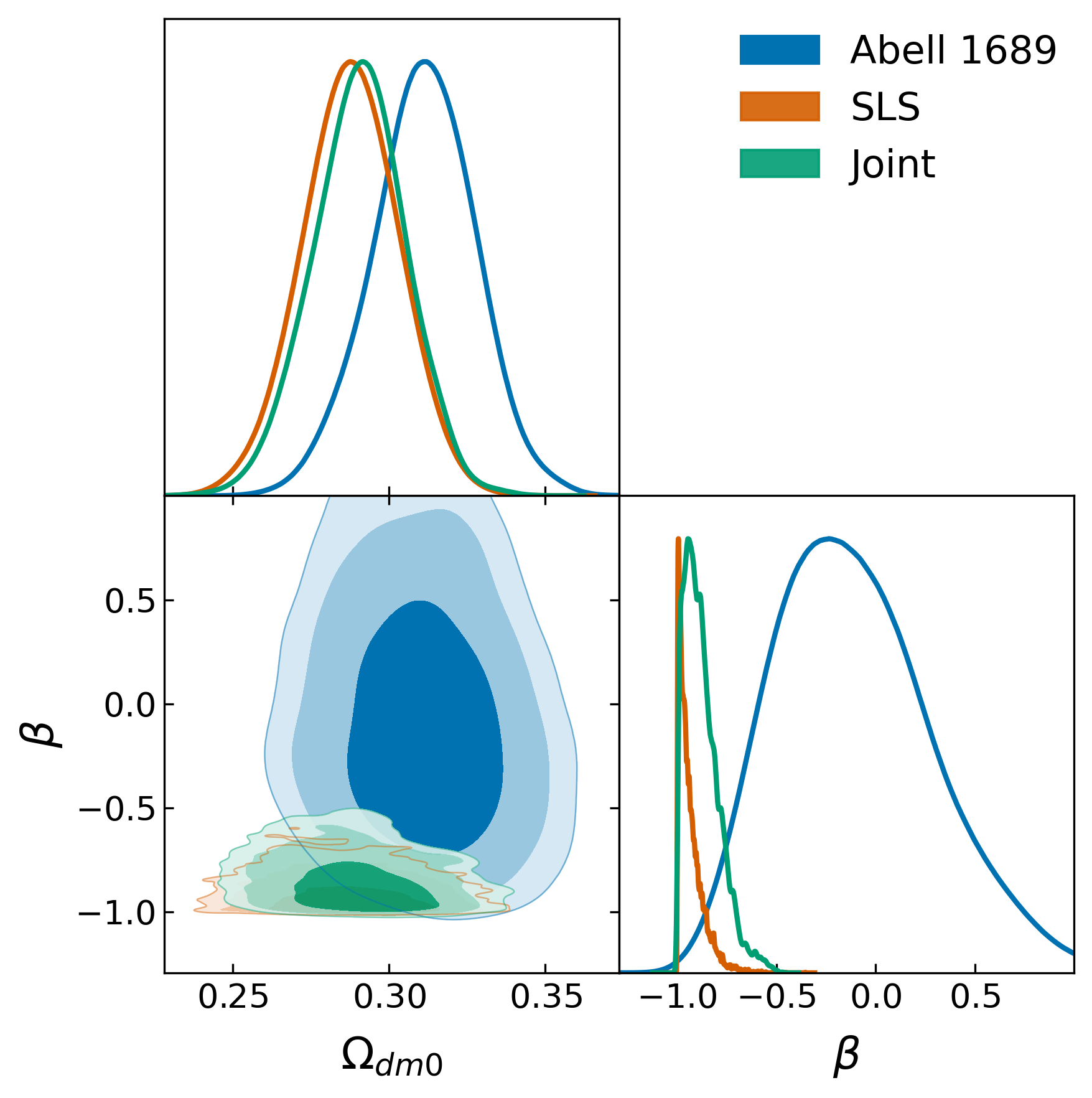}
\caption{Distributions and confidence contours at $1\sigma$, $2\sigma$, and $3\sigma$ for $\Omega_{dm0}$ and $\beta$.}
\label{fig:triangle_plot}
\end{figure}

\begin{figure}[!t]
\centering
\includegraphics[width=\columnwidth]{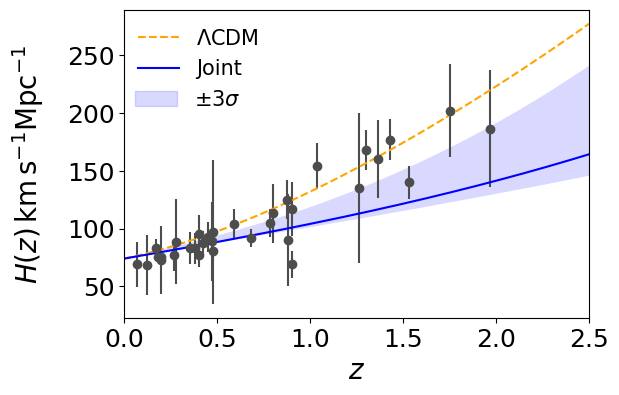}
\caption{Hubble parameter $H(z)$ as a function of redshift. Solid and dashed lines show the interacting and $\Lambda$CDM models, respectively. Points are cosmic chronometer data \citep{moresco_CC_2016}. Shaded region: $3\sigma$ confidence level.}
\label{fig:Hz}
\end{figure}

For our joint analysis (see Fig.~\ref{fig:triangle_plot}), we obtain the best-fit parameters 
$\Omega_{dm0} = 0.253^{+0.018}_{-0.004}$ and 
$\beta = -0.83^{+0.10}_{-0.08}$.
Compared with the results obtained by \citet{wei_interaction_2011}, who found values close to $\Omega_{dm0} \simeq 0.27$ 
and a mildly negative $\beta \sim -0.01$ using supernovae, CMB, and BAO data, 
our strong-lensing constraints favor a larger negative interaction parameter. 
This suggests a stronger coupling within the dark sector than previously reported, 
leading to a more efficient energy transfer from DE to DM
and producing a constantly accelerating Universe.

As shown in Fig.~\ref{fig:triangle_plot}, the cluster analysis exhibits a broad range of solutions along the adopted $\beta$ prior. 
The lower bound of the negative prior has been imposed to avoid divergences in the model, as noted by \citet{wei_interaction_2011}, 
where the interaction term becomes singular for large negative $\beta$ values. 
In this context, the prior interval should be interpreted as a broad exploratory range, while the physically viable region is determined by requiring positive energy densities and an expanding Universe.
The difference between the cluster and galaxy constraints highlights the complementarity of the two strong-lensing methodologies, 
demonstrating that their combination efficiently reduces degeneracies in parameter space. 
This joint behavior reinforces the results previously reported in \citet{verdugo_EPJC_2024}, 
where multi-scale lensing analyses were shown to significantly improve cosmological parameter estimation.

The reconstructed expansion history $H(z)$ (see Fig.~\ref{fig:Hz}) shows that the interacting model 
predicts a systematically lower expansion rate than the standard $\Lambda$CDM scenario at intermediate and high redshifts, 
while remaining consistent with observational $H(z)$ data within the $3\sigma$ confidence level. 
This behavior highlights the effect of the interaction term, which moderates the growth of $H(z)$ with redshift 
and tends to relieve the tension observed between local and high-redshift determinations of the expansion rate \citep{DiValentino2021,Riess2022ApJ}.

\begin{figure}[!t]
\centering
\includegraphics[width=\columnwidth]{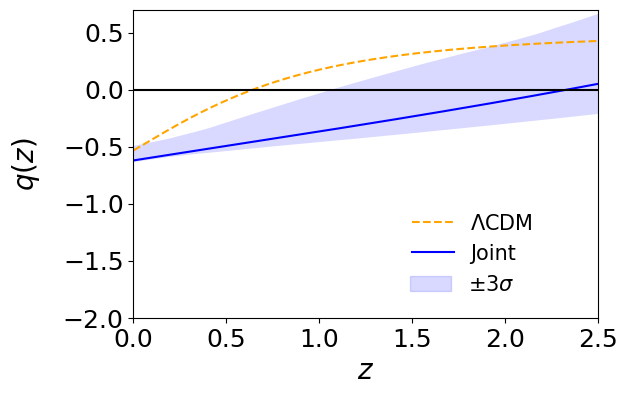}
\caption{Deceleration parameter $q(z)$ as a function of redshift. Solid and dashed lines show the interacting and $\Lambda$CDM models, respectively. The transition redshift $q(z_t) = 0$ marks the onset of cosmic acceleration. Shaded region: $3\sigma$ confidence level.}
\label{fig:qz}
\end{figure}

The deceleration parameter $q(z)$ reconstruction (see Fig.~\ref{fig:qz}) further illustrates the impact of 
the negative interaction parameter obtained in our analysis ($\beta \simeq -0.8$).
Our results show an earlier transition to cosmic acceleration compared with 
the standard $\Lambda$CDM model: the transition redshift, defined by $q(z_t)=0$, occurs at 
$z_t = 2.33$ for the interacting model, in contrast with $z_t = 0.64$ for $\Lambda$CDM \citep{Planck_2018}.  
This earlier onset of acceleration suggests that the dark sector interaction significantly influences 
the timing and dynamics of cosmic acceleration, reinforcing the potential of strong gravitational lensing 
as an effective probe of alternative cosmological scenarios \citep{magana_2015, Magana2018ApJ}.





\section{Conclusions}\label{sec:Conclusions}

We analyzed an interacting $Q$ model using strong lensing data from early-type galaxies and the Abell~1689 cluster following the methodology presented in \citet{verdugo_EPJC_2024}
Our results yield best-fit parameters 
$\Omega_{dm0} = 0.253^{+0.018}_{-0.004}$ and 
$\beta = -0.83^{+0.10}_{-0.08}$, 
favoring a significantly stronger dark-sector coupling than the one reported by \citet{wei_interaction_2011}.
This large negative value of $\beta$ implies a dominant energy transfer from DE to DM, 
producing a faster dilution of the matter component and sustaining a constantly accelerating expansion. 
Such a strong coupling is expected to affect matter perturbations and large-scale structure formation, though a full perturbative analysis is required for a detailed assessment.
The reconstructed expansion history shows a slower growth of $H(z)$ relative to $\Lambda$CDM, 
while remaining consistent with observations within $3\sigma$, and the transition to acceleration occurs earlier, at $z_t = 2.33$.


This joint analysis provides a framework to probe the possible interaction between DM and DE through independent gravitational lensing observations. These results underscore the potential of gravitational lensing as a powerful and independent cosmological probe, consistent with previous studies suggesting that interactions within the dark sector may help alleviate current tensions in the expansion history of the Universe \cite[see][]{Wang2016,Kumar2017}.


\begin{acknowledgement}
F. Villalobos, T. Verdugo, J. Magaña acknowledge support from the Secretaría de Ciencia, Humanidades, Tecnología e Innovación (SECIHTI) under grant CBF-2025-I-551 (“Convocatoria Ciencia Básica y de Frontera 2025”).
FV, JM, PTI, TV acknowledge the support provided by the Agencia Nacional de Investigación y Desarrollo (ANID) through its program Fomento a la Vinculación Internacional (FOVI) 240098. FV and JM acknowledge and the Fondo Nacional de Desarrollo Científico y Tecnológico Fondecyt Regular No. 1240514.
FV acknowledges the use supercomputer of GÜINA (EQM 200216) and support from Universidad Central de Chile from "Facultad de Ingeniería y Arquitectura", "Vicerectoría de Investigación", and the project "Puente de Investigación UCEN 2025".
\end{acknowledgement}


\bibliographystyle{baaa}
\small
\bibliography{bibliografia}
 
\end{document}